\documentclass[aps,pre,a4paper,superscriptaddress,twocolumn,showpacs,showkeys]{revtex4}
\usepackage{bm}
\usepackage{amsmath}
\usepackage{amsfonts}%
\usepackage{amssymb}%
\usepackage{graphicx}
\usepackage{color}

\begin{document}

\title{Performance of the Cell processor for biomolecular simulations}
\author{G. De Fabritiis}
\email[]{gdefabritiis@imim.es}
\affiliation{Computational Biochemistry and Biophysics Lab (GRIB-IMIM/UPF),
Barcelona Biomedical Research Park (PRBB), C/ Dr. Aiguader 88, 08003, Barcelona, Spain. 
 \url{http://www.iac.rm.cnr.it/~gianni}
}

\begin{abstract}
The new Cell processor   represents a turning point for computing intensive
applications.  Here, I show that for molecular dynamics it is possible to reach
an impressive sustained performance in excess of 30 Gflops  with a peak of 45
Gflops for the non-bonded force calculations, over one order of magnitude
faster than a single core standard processor.  
\end{abstract}
\maketitle

\section{Introduction}
The Cell Broadband Engine (CBE)\cite{ibmcellsite} is a new processor
architecture created by Sony-Toshiba-IBM which allows for high computational
performance and low production costs removing  important bottlenecks of
standard processors.  In the present version, it comprises one PowerPC
processing element (PPE) which runs the operating system and acts as a standard
processor and $8$ independent synergetic processing elements (SPEs).  Main
memory can be accessed only by the PPE core while each SPE can use its limited
in-chip local memory (local store) of $256$ Kb accessed directly without any
intermediate caching.  Each core (PPE or SPEs) features a single instruction
multiple data (SIMD) vector unit whose combined peak performance is about 230
Gflops at 3.2 Ghz for single precision floating-point operations. Currently,
double precision units of the SPEs are an order of magnitude slower, a
situation which should improve starting on the next version of the Cell
processor. The main elements of a SPE are a data processing core also called
synergetic processing unit (SPU)  and a memory flow controller (MFC) which
handles communications between main memory and the local memory of the SPE. A
direct memory access (DMA) operation can be initiated by the SPU asynchronously
allowing for overlapping communication and computation and hiding  the cost of
loading data into the local store. The SPU processes  data available on its
local store removing the memory  bottleneck which is afflicting modern
processors. Finally, on a dual processor blade, a program can request all 16
SPEs transparently. 

All this computational power comes at the  cost  of a programming paradigm
change which requires using  multi-threading and vectorized code.  The Cell
processor can be programmed  as a multi-core chip with nine heterogeneous cores
using standard ANSI C and relying on the libraries from the IBM system development kit
(SDK) to handle communication, syncronization and SIMD computation.
Programmability is an important aspect which distinguishes the Cell processor
from other specialized processors, e.g. graphical processing units (GPUs). The
Cell processor requires a set of advanced but standard programming techniques
which are already in use on standard multi-processor machines  supporting common
programming constructs and languages like C/C++. The overall performance is
strongly dependent on the  the effective use of Cell hardware  which is largely
left to the code and compiler.  However, each step in the optimization can be
taken incrementally. An existing application would run on the Cell processor by
a simple recompilation of the code using only the PPE core, with no effort, but
also without advantages from a performance viewpoint.  In order to obtain the
{\it highest} performance,  it is necessary to use all the SPEs, vector
hardware and to adapt to the memory access architecture.  

Vectorization of the code is very important because  the SPEs are not optimized
to run scalar code and handling unaligned data. A SIMD add instruction
(spu\_add) allows to compute four floating-point add operations in a single
instruction operating on a $128$ bits type (vector float) that is the
combination of four floats (code samples are found in the CBE
tutorial\cite{CBEtutorial}).  These intrinsic primitives are for the most part
derived from the more standard AltiVec instruction calls in the PowerPC element
(vec\_add).  The compiler automatically aligns vector types to $16$ bytes
memory boundaries which can then be loaded  directly into the SPE registers.
Manual data alignment and padding are also necessary  for  data communications
between local stores and main memory. 

After vectorization of the computing intensive parts of the code, the work must
be distributed on multiple SPEs using multi-thread programming techniques which
entails handling  synchronization between processing threads running on the
$9$ processing cores of the Cell processor. The libraries of the SDK provide
several ways to control SPE threads which in most cases are similar to other
multi-thread libraries.  It is also best to avoid conditional branching in the
computational intensive parts of the code because SPEs lack appropriate
hardware for branch prediction.  

Optimizations discussed so far would be beneficial to standard processors as
well (for instance using the streaming SIMD extensions (SSE)). Specific to the
Cell processor is the SPE  core design which makes {\it all} these optimization
steps  crucial for performance and the local store which provides very fast
access to local data. The SPE core design   provides reduced power consumption
and higher clock frequencies, while the memory architecture is designed to
avoid that the fast synergetic processing units are ever starving for data
coming from the slow main memory.  This new memory architecture requires the
programmer to think of algorithms which fit in the limited $256$ Kb of the
local store of each SPE and the communication between local store and main
memory using DMA  calls of the system development kit.  Overall, good knowledge
of standard parallel and vector programming techniques represents the largest
learning obstacle to program the Cell processor, as well as standard multi-core
chips. 

In the following, first results of biomolecular simulations are presented 
 supporting the idea that the  Cell processor has the potential for being
interesting   for a widespread set of applications thanks to a nice programming
interface (C plus libraries), low cost per chip and sustainable performance
which makes a difference.

\begin{figure}[bt!]
\begin{center}
\includegraphics[width=6.5cm]{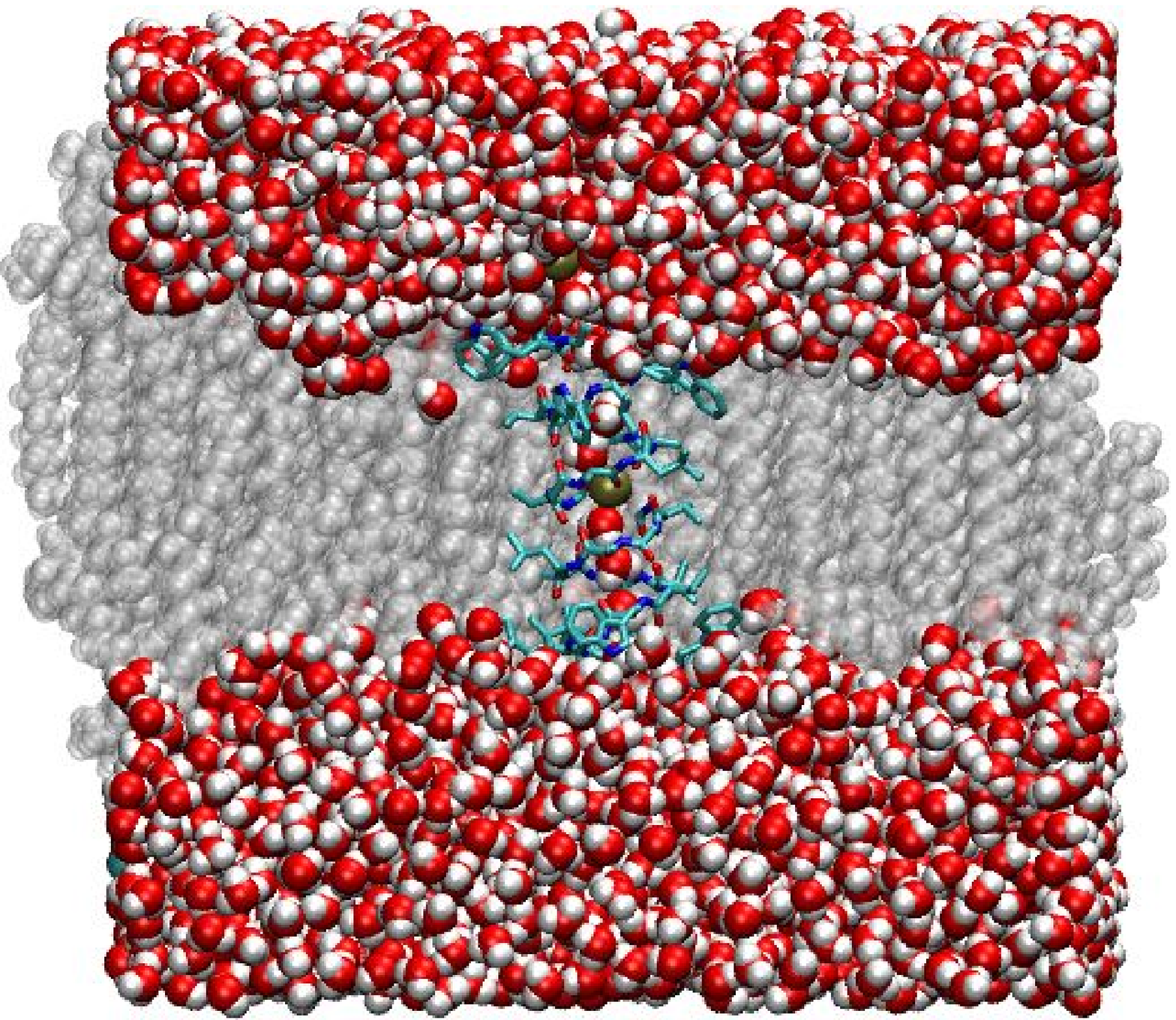}
\end{center}
\caption{Ion channels are essential to life by regulating transport across
lipid membranes.  Here, it is shown a Gramicidin A channel solvated in a lipid
membrane with water and ions. Lipids are only partially shown for
clarity\cite{VMD}.  } 
\label{gramafig}
\end{figure}

\section{Molecular dynamics case study}
Molecular dynamics (MD) is a simulation methodology which enables, for
instance, the study of the dynamics of proteins in their environment. It is
used by pharmaceutical companies for a wide variety of applications including
drug design, drug screening  and, in general, to investigate protein function.
This has been achieved through the use of  carefully tuned force fields which
reproduce the molecular specificity of each protein\cite{charmm27}. However,
the impact of molecular dynamics would be much greater if faster ways to
perform MD simulations were found in order to reach the time scales of
biological processes (micro-milli seconds).  These time scales cannot be
simulated yet despite the use of costly high performance supercomputers with
thousands of processors.  Specialized hardware like the Cell processor could
help to approach this goal.

The molecular dynamics software presented in this paper for benchmarking the
performance of the Cell processor is able to read CHARMM27 force
fields\cite{charmm27} and to simulate bio-molecular models such as proteins,
lipids and TIP3P water with periodic boundary conditions.  Electrostatic and
Lennard-Jones interactions are handled by  simple truncation with switching
functions used to smooth the force to zero at  the cutoff radius. A  cell index
method is used to handle non-bonded interactions within the cutoff
radius\cite{allen87}. This  code provides already a functional MD engine for
bio-molecular simulations which is being used  for applications such as ion
permeation of protein channels\cite{defabritiis_ga}.  For the current
benchmarks, the MD simulations are run on the molecular system depicted in Fig.
\ref{gramafig} which consists of  Gramicidin A trans-membrane protein embedded
in a DMPC lipid bilayer and water for a total of $29$ thousand
atoms\cite{defabritiis_ga} and on TIP3P water boxes of different sizes. Elapsed
time is measured over at least $500$ iterations for short runs in order to
remove the cost to start-up the simualtion and then rescaled to $50$
iterations.  The cutoff radius is set to $12$ {\AA}, switching distance $10$
{\AA} and pair interactions are updated every $20$ iterations (a cell size of
$13.5$ {\AA} is used to account for diffusion of atoms during this time lapse).
The architectures used for the benchmark are 1) an Opteron based PC at 2Ghz
with Linux Fedora Core 5 and gcc compiler version 4.1.1 and 2) Fedora Core 5
with the IBM system development kit  version 1.1 for the Cell blade running at
3.2 Ghz. 

\begin{figure}[t!]
\begin{center}
\includegraphics[width=7.5cm]{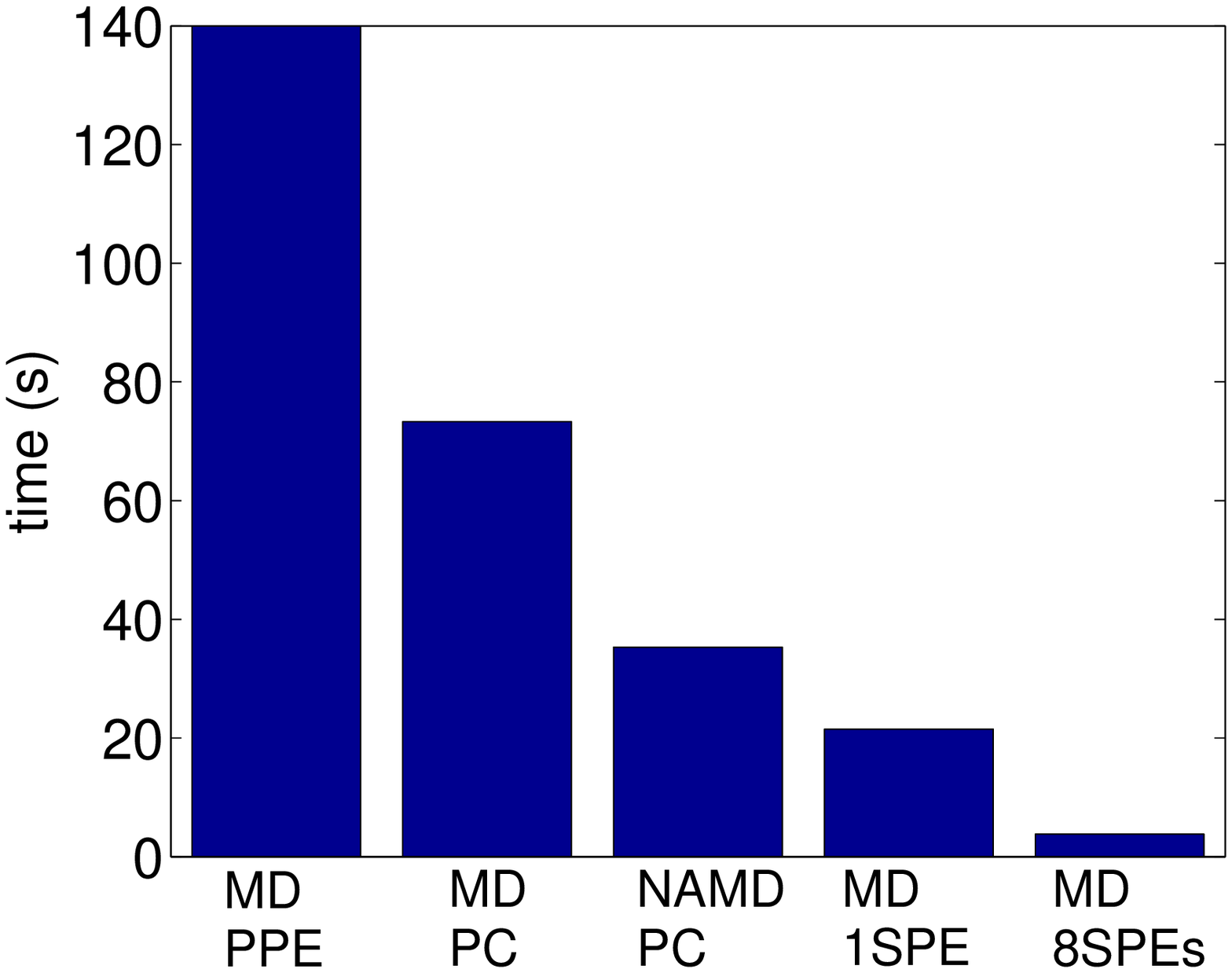}
\end{center}
\caption{Simulation speed-up to run $50$ iterations of Gramicidin A in the set-up
of Figure \ref{gramafig}. The MD code is run on the PPE, on a PC Opteron and on
$1$ and $8$ SPEs.  The Cell MD code is over $19$ times faster than the scalar code. A
speed-up of $35$ times is obtainable by running on $8$ SPEs compared to the PPE. }
\label{benchfig1} \end{figure}

As a first benchmark, the MD code  is compared with a widely used molecular
dynamics package NAMD2.6 \cite{NAMD}, specially optimized for parallel
processing, but also very fast on a single processor.  NAMD results twice as
fast as the scalar MD code  running on the Opteron processor  using equivalent
input parameters.  This performance difference is due to algorithmic
optimizations  of modern MD engines (table look-up for potentials and faster
pairlist creation). Nevertheless, with the current implementation, the  Cell MD
code on $1$ SPE is already faster than NAMD on the Opteron processor and up to
an order of magnitude faster on $8$ SPEs (Figure \ref{benchfig1}). As such,
this code is sufficiently fast to represent a significant benchmark for the
performance of the Cell processor for biomolecular simulations. 

The MD code for the Cell processor shares the same algorithmic solutions as the
scalar MD code. However, the SPEs are optimized for single precision
floating-point operations which were then used for non-bonded calculations. Single
precision calculations are reasonable for molecular dynamics because the
trajectory is chaotic and additional digits of precision are quickly lost in
few hundred iterations. The correctness of the single-precision floating point
implementation is tested by comparing with the single-precision scalar
implementation directly derived by the double precision code. The double
precision results  are also compared with NAMD. For single precision, scalar
and vector units produce negligible differences in the total energy due to
different round-off errors. Furthermore, simulations are run in the NVE
ensemble which allows  to check that the total energy remains approximately
constant over time, oscillating around the mean value. 

The Cell MD code runs on the PPE for all parts except for the calculations of
the non-bonded forces.  The code running on the PPE still uses double precision
floating-point scalar operations as the original code, while the calculations
of Lennard-Jones and electrostatic forces and potentials on the SPEs are coded
using single precision SIMD vector instructions.  Atom positions are buffered
to single-precision values before being sent to the local store of the SPE. The
forces and energies computed on the SPE are converted again to double precision
when returned to the PPE.  Work distribution is handled by the PPE thread
running  the program and managing the SPE threads in a master-slave protocol
using mailboxes to communicate and to synchronize threads.  The PPE assigns
iteratively a set of  non-bonded interactions   to the SPEs  which  return the
computed forces and energies. All the code is written in plain ANSI C  using
only the IBM SDK libraries which provide intrinsic C calls for managing  SIMD
units, mailboxes and SPE threads (a detailed description can be found in the
IBM Cell tutorial\cite{CBEtutorial}).  Thanks to these primitives provided by
the SDK there is no need to use assembly language with the result of
simplifying a lot the work of the programmer but still achieving impressive
sustained performance.

\begin{table}[tb]
\caption{ Elapsed time to run 50 iterations of Gramicidin A.}
\begin{tabular}{|c|c|c|c|c|c|c|}
\hline
 Time (s) & 1 SPE & 2 SPEs & 4 SPEs& 6 SPEs  & 8 SPEs & 16SPEs\\
\hline
 Total & 21.5 & 11.1 & 6.3 &  5.1 & 3.8 & 3.5\\
 Speed-up & 1 & 1.9 & 3.4 &  4.2 & 5.6 & 6.1\\
\hline
\end{tabular}
\label{elapsed_table}
\end{table}


In Figure \ref{benchfig1},  the first two columns show the scalar code running
on the PPE of the Cell processor and on the Opteron processor. The first
expected result is that the PPE is outperformed by an Opteron chip although it
runs at a much higher clock frequency. This is due to fact that the PPE is not a
fully equipped PowerPC but rather a simplified version designed to reduce power
consumption and leave space on die for the SPEs on which computing intensive
tasks are expected to run. As a matter of fact, the Cell MD code with just $1$ SPE is
already $6$ times faster than the scalar version running on the PPE.

In Table \ref{elapsed_table}, the performance speed-up  is shown for $1$ to
$16$ SPEs.   A loss of efficiency  between $1$ to $8$ SPEs
is partially due to the fact that non-bonded force calculations, which have been
parallelized across the SPEs, become comparable to other parts of the force
calculations which are still running on the PPE like bonded terms and geometric
hashing using the cell index method. A faster PPE would directly improve the
performance of the entire code. However, the current implementation seems also
to suffer particularly on $16$ SPEs, i.e. using both processors on the blade.
This is being investigated and hopefully will be solved in a future version,
but it should not be a problem of bandwidth saturation of the interconnection
bus\cite{petrini06}.  For $8$ SPEs, corresponding to one single Cell processor,
the calculated sustained performance of Cell MD is in excess of $30$ Gflops.
This increases to $45$ Gflops if we consider only the non-bonded force
calculation, which is very good sustained performance. Flops are measured by counting
the number of vector single-precision floating point operations and
distinguishing between vector multiply-add operations ($8$ flop) and more
common simple vector multiplies ($4$ flop). The inner loop of the force
calculation performs at an impressive $0.7$ clocks per instruction (CPI) which
compares to a theoretical minimum of $0.5$ CPI, achieved when odd and even
pipelines on the SPEs can both issue an instruction each clock.   

\begin{figure}[tb]
\begin{center}
\includegraphics[width=7cm]{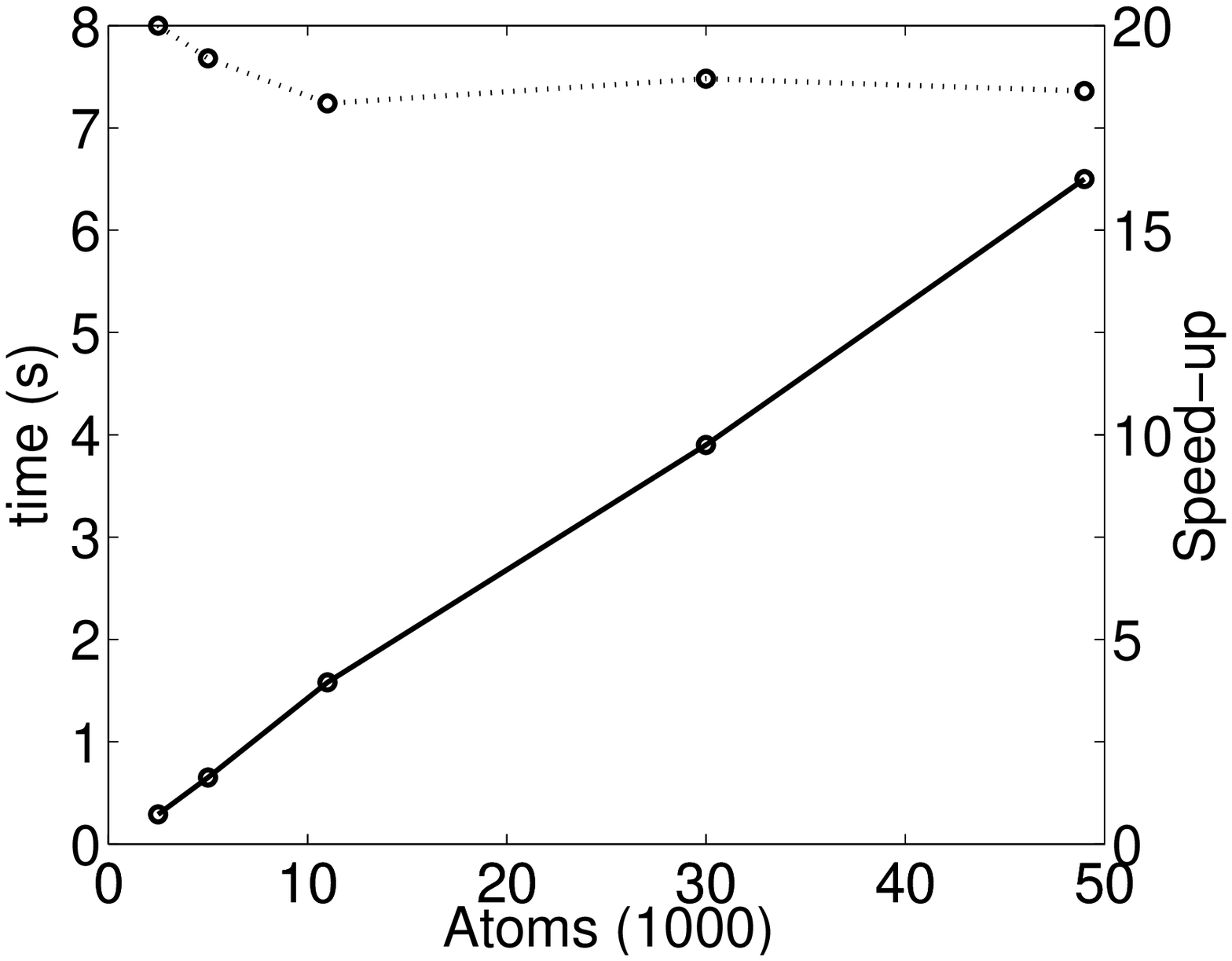}
\end{center}
\begin{tabular}{cccccc}
 Atoms & 2.5K & 5K & 11K &  30K& 49K\\
\hline
 Opteron PC & 5.8 s & 12.5 s & 28.6 s& 72.9 s& 119.6 s\\
 Cell 8 SPEs & 0.29 s & 0.65 s& 1.58 s& 3.9 s&  6.5 s\\
\end{tabular}
\caption{ Elapsed time (circles) over $50$ iterations for Cell MD on 8 SPEs
for water boxes at different problem sizes: $2.5$, $5$, $11$, $30$ and $49$ thousand
atoms. The scaling is linearly dependent on problem size. The speed-up  (dotted
line) compared to the scalar code run on the Opteron processor is reported on
the right axes. }  \label{benchfig3} 
\end{figure}

Another important factor is the scalability of the Cell MD code for varying
number of atoms. Better performance can be achieved by a further domain
decomposition of the code running on multiple Cell hardware connected by a fast
network. In this case the  atoms would be partitioned across the processors to
achieve higher speed by reducing the number of atoms per processor. It is
therefore crucial that the scalability of the code on number of atoms remains
optimal (linear on problem size) even for the smallest system. Figure
\ref{benchfig3} shows a benchmark of the elapsed time for $50$ iterations of
water boxes with $2.5$, $5$, $19$, $30$, $49$ thousand atoms all running on $8$ SPEs.
The scaling is linearly dependent on the system size, therefore there is no
loss of performance  even for the smallest system. A final evaluation of
the parallel performance would of course require to test the parallel
implementation itself. Nevertheless, this result indicates that a low number of
atoms per Cell processor should be reachable maintaining optimal efficiency.
In this case, a scaling to six dual processor blades for gramacidin A  corresponds
to approximately 120 standard cores which would be a very good result if a
parallel MPI Cell code was to confirm it.
On the other hand, the largest system which can be
simulated depends essentially on the main memory available on the system. 
º

\section{ Conclusion} 
In conclusion, the Cell processor runs existing applications on the standard
PowerPC core  but in general a performance penalty should be expected compared
to other ``off-the-shelf'' processors. The strength of the Cell processor is 
the synergetic processing units which require advanced programming skills such
as SIMD vectorization and knowledge of parallel and multi-threaded programming
together with a good understanding of the architecture of the Cell processor.
The cost of this effort cannot be underestimated. A quantitative estimation of
the time required to produce good Cell code crucially depends  on  previous
knowledge of the programmer/scientist on these advanced programming techniques
and on the application domain. In the best case, it should be considered as a
standard parallelization task which may take from weeks to months if maximum
performance is required.  All the development can be done in standard ANSI C
with the use of special libraries. Also, the application code plays an
important factor: molecular dynamics is dominated by the cost of non-bonded
force calculation which is very computing intensive. 

The performance  obtainable compared to a traditional processor is about $20$
times faster for the realistic case of molecular dynamics of biomolecules which
easily justifies the effort  for this computational demanding application domain.
Similar results are also possible for other computing intensive scientific and
technological problems \cite{dongarra2006cell, yelick06} such as computational
fluid dynamics, systems biology and Monte Carlo methods for finance. We plan to
extend this work to these applications  in the very near future.  The
performance measures of this article are to be considered conservative but
quite accurate. Optimizations are  in progress which could further enhance  the
speed of the Cell MD code by achieving a better scaling of the code on multiple
SPEs. 

  The innovative design and low cost per chip of the Cell processor are likely
to be  key factors in the probable success of this type of technology. Part of
the cost benefits comes from the fact that the  Sony PlayStation3 \cite{ps3}
features the Cell processor guaranteeing high production volumes from the very
beginning. Standard multi-core  processors will need to show that they can
reach similar performance levels at the same cost. In the future, it will be
interesting to benchmark also an SSE optimized version of this code on standard
multi-core processors. However, two are the key important advantages of the Cell
processor. The simplified SPE cores allow for reduced power consumption and
space occupation on die which make possible to put more cores on the chip. The
memory architecture of the Cell processor resolves the memory bottleneck which
afflicts  multi-core standard processors specially when number of cores starts
to become significant. Because of this, the Cell BE architecture provides a
possible scalable technology which could allow within a decade to
reach  routinely millisecond time scales in molecular simulations. New
non-standard processor technologies from Intel (specialized $80$ floating-point
processing cores) and the stream computing initiative from AMD-ATI are both
exciting additions  which already prove the impact that the Cell processor has
had.  The implications of this technology for  science are very important.
Without a doubt it expands the frontier of scientific computing while lowering
the cost of entry in terms of the computational infrastructure required to run
molecular based software. 

\section*{Acknowledgments.}
I am grateful to Barcelona Supercomputing Center (BSC) for support and
access to Cell hardware  and to Giovanni Giupponi, David Hardy, Matt Harvey,
Massimo Bernaschi, Peter Coveney and  Jordi Vill\`{a}-Freixa for their help.
\bibliographystyle{apsrev} 
\bibliography{../gianni}
\end{document}